\documentclass{article}  
\usepackage{abstract_2e} 

\usepackage{times}
\usepackage{epsfig}

\begin{document}  

\title{Exploring the Inner Edge of the Habitable Zone with Fully Coupled Oceans}

\author{M. J. Way}
\affil{NASA Goddard Institute for Space Studies, New York, NY, USA (michael.way@nasa.gov)}
\affil{Department of Astronomy and Space Physics, Uppsala, Sweden}
\author{A. D. Del Genio}
\author{M. Kelley}
\author{I. Aleinov}
\affil{NASA Goddard Institute for Space Studies}
\author{T. Clune}
\affil{NASA Goddard Space Flight Center, Greenbelt, Maryland, USA}


\runningtitle{}

\titlemake  

\begin{abstracttext}
Rotation in planetary atmospheres plays an important role in regulating
atmospheric and oceanic heat flow, cloud formation and precipitation. Using the
Goddard Institute for Space Studies (GISS) three dimension General Circulation
Model (3D-GCM) we demonstrate how varying rotation rate and increasing the incident
solar flux on a planet are related to each other and may allow the inner edge of
the habitable zone to be much closer than many previous habitable zone studies
have indicated.  This is shown in particular for fully coupled ocean runs -- some
of the first that have been utilized in this context. Results with a slab ocean 100m mixed
layer depth and our fully coupled ocean runs are compared with those of
Yang et al.\cite{Yang2014}, which demonstrates consistency across models using a slab ocean.
However, there are clear differences for rotations rates of 1-16x present Earth
sidereal day lengths between the 100m mixed layer and fully couple ocean models,
which points to the necessity of using fully coupled oceans whenever possible.
The latter was recently demonstrated
quite clearly by Hu \& Yang\cite{HY2014} in their aquaworld study with a fully coupled ocean
when compared with similar mixed layer ocean studies and by Cullum et al.\cite{Cullum2014}.

Atmospheric constituent amounts were also varied alongside adjustments to cloud
parameterizations. While the latter have an effect on
what a planet's global mean temperature is once the oceans reach equilibrium they
do not qualitatively change the overall relationship between the globally averaged
surface temperature and incident solar flux for rotation rates studied herein.
At the same time this study demonstrates that given the lack of knowledge about the
atmospheric constituents and clouds on exoplanets there is still a large uncertainty
as to where the boundaries of a planet's habitable zone exist for a given star.
We also show how these results have implications for Venus in the early history
of our Solar System.

\section*{Methodology and Model Inputs}

We have made a number of modifications to the default modern-day Earth model for
this study to better mimic the Yang et al.\cite{Yang2014} model setup.

\begin{enumerate}

\item Ocean 1 (Fully Coupled): A fully coupled ocean with present day
temperature, depth and salinity are used At Model Start (AMS).

\item Ocean 2 (Slab): a 100 meter depth Q-Flux\cite{russell1985} ocean, but with
the horizontal fluxes set to zero (Qflux=0). Yang et al.\cite{Yang2014} use a 50m Q-flux=0 ocean.

\item Atmospheric constituents 1: N$_{2}$=980mb,\newline
CO$_{2}$=0.392bar (400ppm), CH$_{4}$=0.00098bar (1ppm),
N$_{2}$O=.00026 (0.27ppm), O$_{3}$=PAL (Present Atmospheric Level Earth), H$_{2}$O=PAL.

\item Ice Cloud Parameterization: Increased ice cloud content at high altitudes with lower
latitudes \emph{and} at lower altitudes with higher latitudes (see Figure \ref{plot3}). This will
tend to increase the long-wave cloud radiative forcing and will warm the planet slightly.

\item Land parameters: Earth Continental Layout. All land albedos are set to 0.2
AMS. There is no land ice AMS. The soil is a 50/50 mix of sand/clay
and there is no vegetation.

\item Orbital Parameters: Both obliquity and eccentricity are set to zero.

\item Model resolution: Latitude-Longitude grid 4x5 degrees in resolution, with
20 vertical layers.

\item Rotation periods chosen: Sidereal (Solar) day length in present Earth days:
1x (1x), 16x (16.7x), 64x (76.6x), 128x (191x), 256x (848x).

\item Solar insolations as increases from present day Earth: 1, 10, 20, 30, 40, 50,
60\%.

\end{enumerate}

Regarding the use of a Slab ocean (item 2), many other studies do not know what
the ocean horizontal heating fluxes are when the rotation rate and/or solar
insolation are changed from present day Earth values. The only way to obtain
these numbers would be via Fully Coupled Ocean runs, but these are time
consuming and computationally expensive.  Hence the fluxes are set to zero in
the belief that this is superior to using the present day Earth fluxes.

The ice cloud parameterization modification (item 4) allowed our 1 Earth day with
present day solar constant model global mean surface temperature (289K) to be closer
to the Yang et al. starting point of 287K. Otherwise our starting temperature
would have been around 285K.

Note that in the upper right hand panel of Figure \ref{plot1} we show the effect
that cloud tuning can have. The dotted gray line (1Bx) is a modern Earth rotation
rate with increasing solar insolation, but this alternative cloud tuning starts
at a much lower initial surface temperature for present day solar insolation,
and hence all increases in solar insolation have resulting lower globally
averaged surface temperatures. This allows one to reach higher solar
insolations before the radiation scheme reaches its limits.

Not all of the higher solar insolations (item 9) could be reached for a
given rotation period because the model radiation tables are only valid to 373K,
but in practice a given grid cell's surface temperature cannot exceed
approximately 350K.

\section*{Figures}
\begin{figure}[!htb]
\includegraphics[scale=0.215]{plotTSURF_E21eoDOFP3Od_E1oM20B.eps}
\includegraphics[scale=0.215]{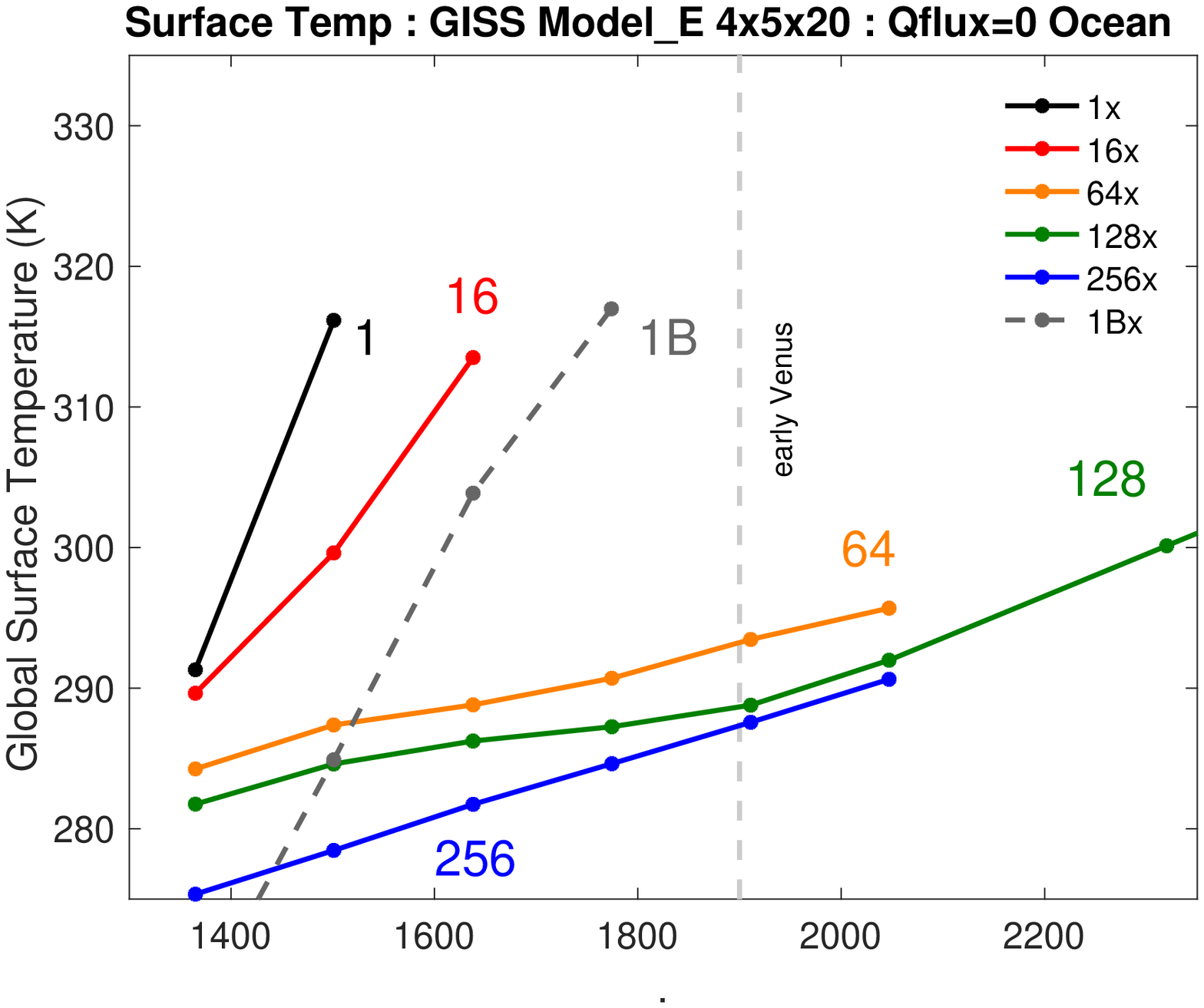}
\includegraphics[scale=0.215]{plotOICEFR_E21eoDOFP3Od_E1oM20B.eps}
\hspace{0.1cm}
\includegraphics[scale=0.215]{plotOICEFR_E21eoZohtCO2400_E4M20B.eps}
\includegraphics[scale=0.215]{plotPLANALB_E21eoDOFP3Od_E1oM20B.eps}
\hspace{0.1cm}
\includegraphics[scale=0.215]{plotPLANALB_E21eoZohtCO2400_E4M20B.eps}
\caption{\small Fully Coupled Ocean left panels, Qflux right.}
\label{plot1}
\end{figure}

\begin{figure}[!htb]
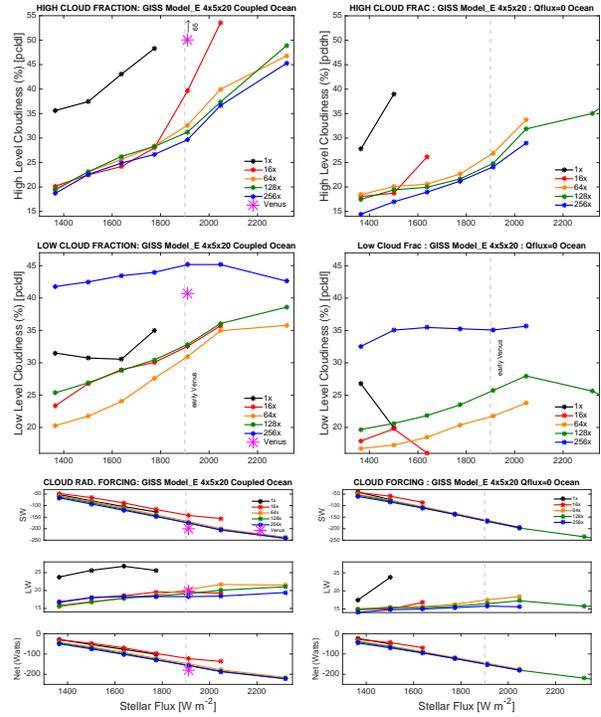

\includegraphics[scale=0.215]{plotPCLDH_E21eoDOFP3Od_E1oM20B.eps}
\hspace{0.1cm}
\includegraphics[scale=0.215]{plotPCLDH_E21eoZohtCO2400_E4M20B.eps}
\includegraphics[scale=0.215]{plotPCLDL_E21eoDOFP3Od_E1oM20B.eps}
\hspace{0.1cm}
\includegraphics[scale=0.215]{plotPCLDL_E21eoZohtCO2400_E4M20B.eps}
\includegraphics[scale=0.21]{plotCRF_E21eoDOFP3Od_E1oM20B.eps}
\hspace{0.1cm}
\includegraphics[scale=0.21]{plotCRF_E21eoZohtCO2400_E4M20B.eps}
\caption{\small Fully Coupled Ocean left panels, Qflux right. LW=Long Wave, SW=Short Wave, Net=SW+LW.}
\label{plot2}
\end{figure}

\begin{figure}[!htb]
\includegraphics[scale=0.35]{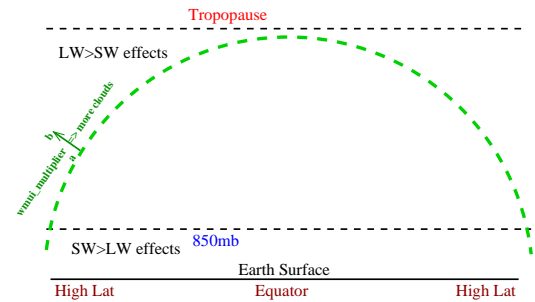}
\caption{\small High altitude ice cloud parameterization change from present day Earth
(case a) to this study (case b). The increased ice cloud content from a-to-b will
tend to increase the long-wave cloud radiative forcing and will warm the planet slightly.}
\label{plot3}
\end{figure}

\begin{figure}[!htb]
\includegraphics[scale=0.28]{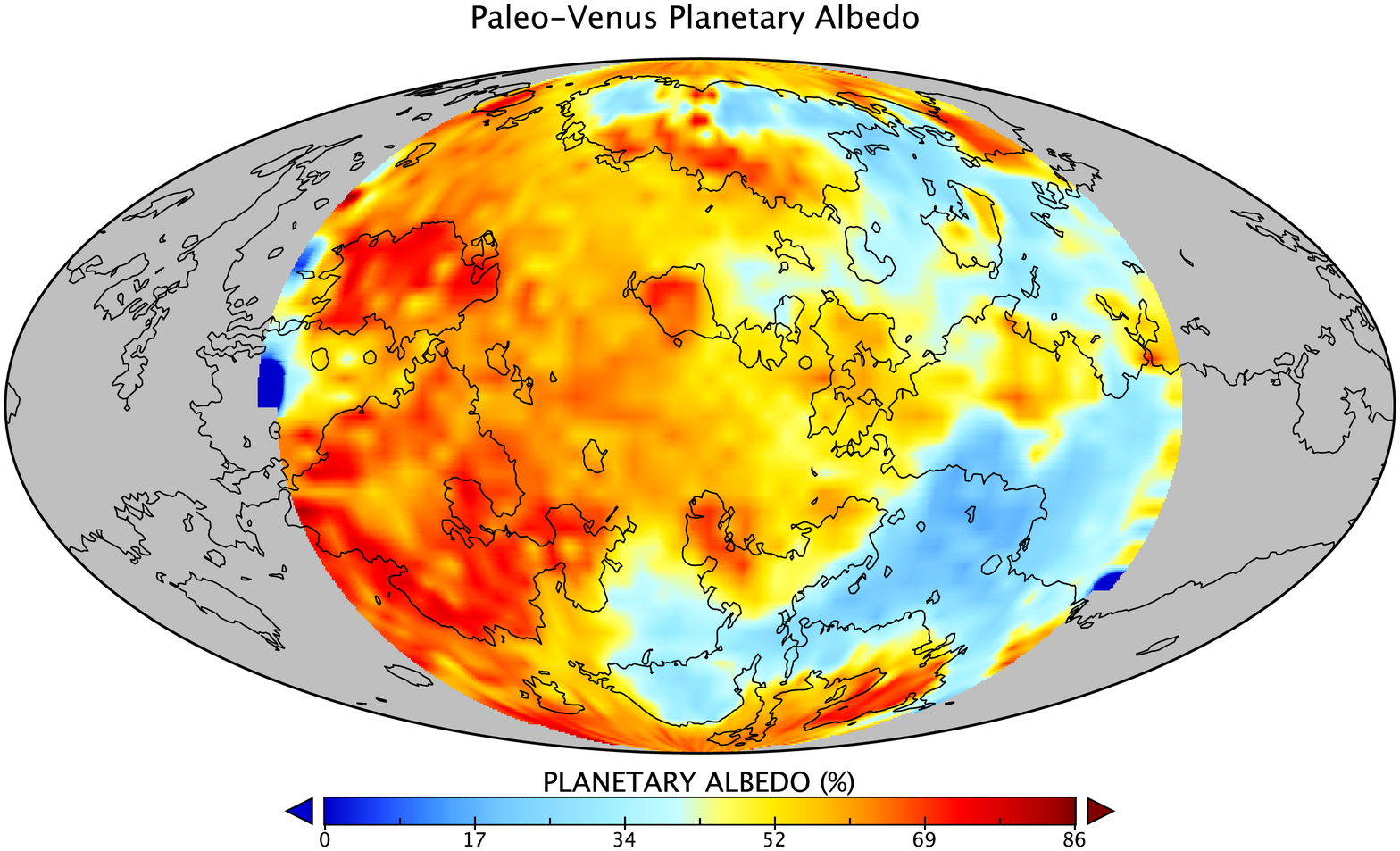}
\includegraphics[scale=0.28]{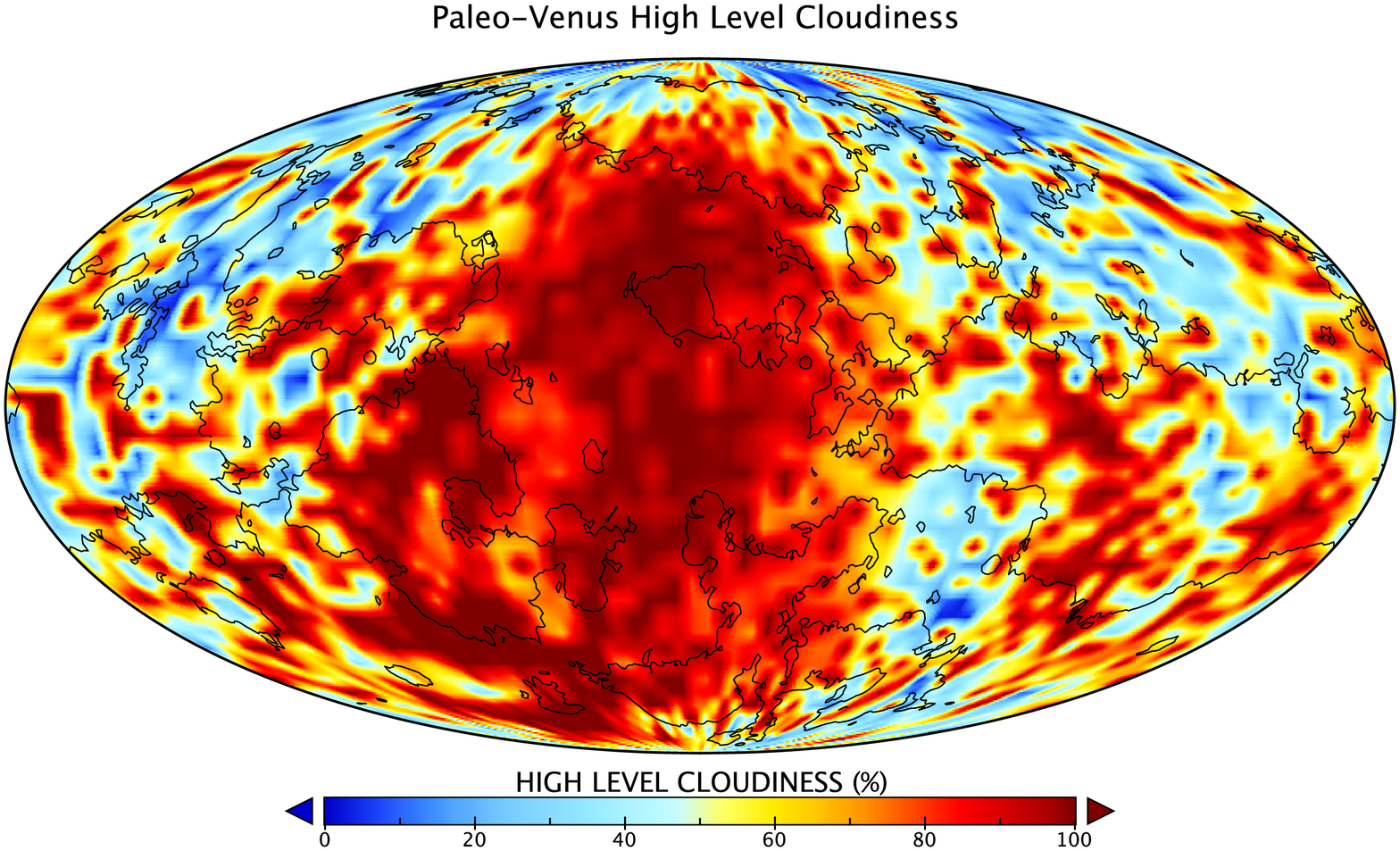}
\includegraphics[scale=0.28]{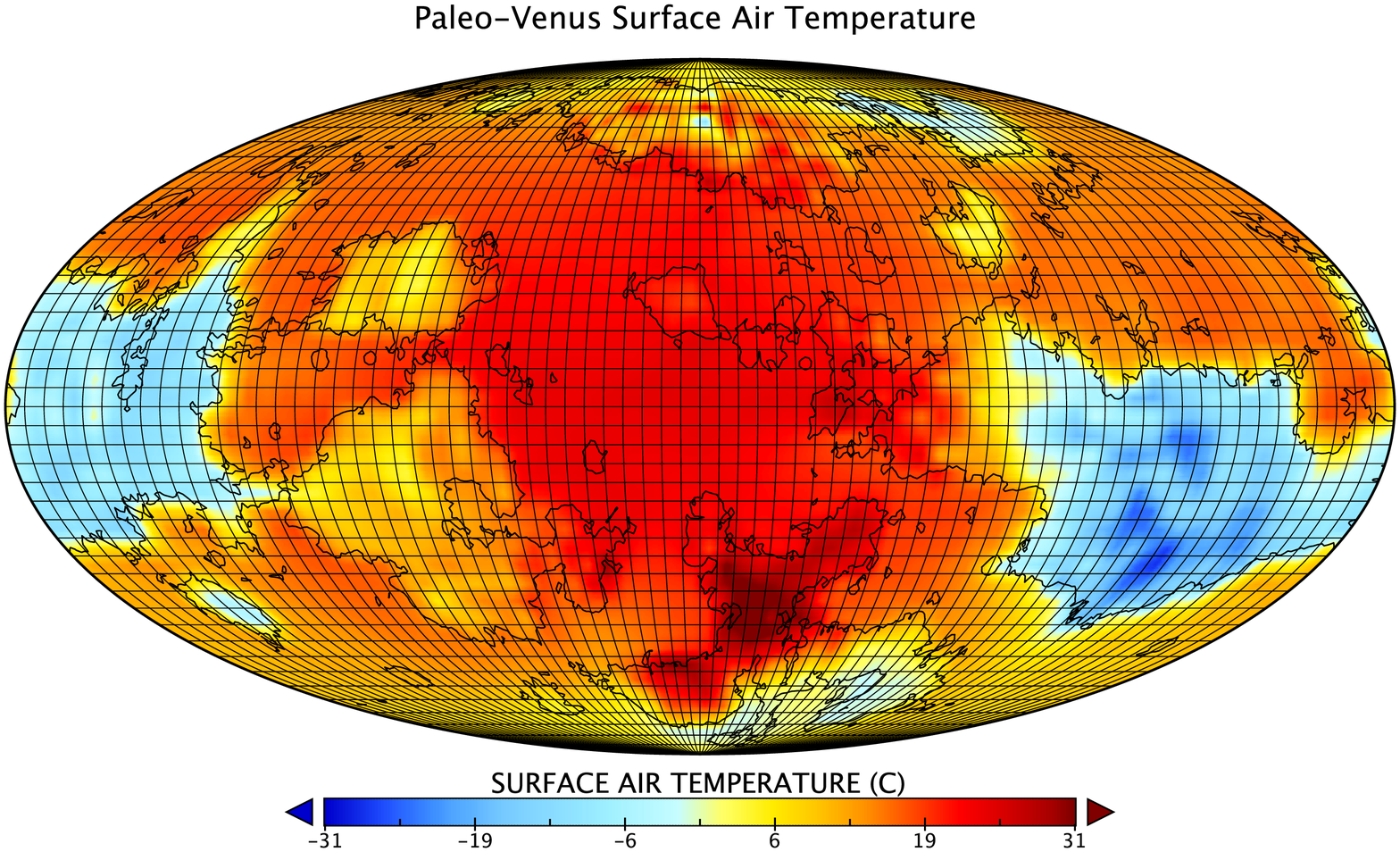}
\caption{\small Top: Planetary albedo for paleo-Venus. Middle: High Level Cloud
Fraction. Bottom: Surface Air Temperature (grid box size is equal to the
model resolution). The figures are snapshot averages over
approximately 1/12 of a Venusian sidereal day.  Notice that the highest value large
concentration of clouds sits at the substellar point.}
\label{plot4}
\end{figure}

\section*{Discussion}

From the surface temperature plots in the top panels of Figure \ref{plot1} it
is clear there is a split in the dynamics between 1--16x present sidereal day
Earth length and 64x--256x regardless of the type of ocean used. Of course much
can happen between 1x--16x (see Wolf \& Toon 2015\cite{WT2015}).
For most solar insolations the ocean ice fractions are different when comparing
the Fully Coupled (left) versus slab ocean (right) (Figure \ref{plot1}).
These types of differences have been observed previously:
compare \cite{Edson2011} and \cite{HY2014}. Planetary albedos
(bottom panels in Figure \ref{plot1}) are also different depending on the
ocean used, but especially in the 1x and 16x rotation rate cases.

We are still investigating why the fully coupled oceans have more high and low
cloud fractions (top 2 left panels in Figure \ref{plot2}) compared with the Qflux=0 oceans.
However, the Cloud Radiative Forcings (bottom 3 panels) in the short wave (SW) are
not so different regardless of the ocean used, while long-wave (LW) forcings appear
to be smaller in the Q-flux=0 ocean cases. The cloud fraction and
thickness of high and low levels in the troposphere may greatly influence the
radiative balance of the planet and hence its temperature.
The water vapor content in most of our cases is well
below the classical water loss limit and hence with future high temperature
extensions to our radiation code we expect to find that planets may be found at
the inner edge of the habitable zone akin to a paleo Venus world (see description further down)
with Solar insolation (S0X) of 1.4 (40\% greater than present day Earth) even with
present day Earth rotation rates (see upper left panel of Figure \ref{plot1}).
We believe the large temperature transition at higher S0X between rotation periods of 16x
and 64x Earth days are likely due to circulation changes being dominated by the
Hadley cells (1x-16x) in contrast to day-night transitions (64x and higher). We believe
this is because the radiative relaxation timescale starts to be less than the
rotation rate timescale at 64x and this fact causes a change in circulation
patterns from x16$\rightarrow$x64. This is also shown in the greater high cloud fraction 
in the Hadley regime (1-16x rotation) while in the day-night regime (64x and
higher) the high cloud fraction is less.

Finally, in the left hand plots (Fully Coupled Ocean) in Figures \ref{plot1}
\& \ref{plot2} we have designated paleo-Venus runs with a purple asterisk. The Venus runs
utilize a 2.9Ga solar spectrum generated with the code of Claire et al.\cite{Claire2012}
(0.5--2397.5nm from Thuillier et al.\cite{Th1,Th2}, and 2397.5nm and higher from
\cite{Claire2012}), a modern Venus topography with an ocean filling the lowlands
(giving an equivalent depth of 310 meters if spread across the entire surface),
atmosphere of 1 bar N$_{2}$O, CO$_{2}$=0.4mb, CH$_{4}$=0.001mb and present day
orbital parameters (spin, obliquity, etc.) radius, and gravity.
Figure \ref{plot4} contains snapshot averages over approximately 1/12 of a
Venusian sidereal day.  Clearly ancient Venus sits within the slowly rotating world ranges
of most of the quantities in Figure \ref{plot1}. This is likely because the
retrograde spin rate of Venus is not too distinct from that of
the 256x case in terms of atmospheric dynamics. The high albedo
(Figure \ref{plot4} top) clearly moderates the surface temperature. This high albedo comes
from large cloud convection at the substellar point (Figure \ref{plot4} bottom),
likely due to the large day-night circulation that comes with a slowly rotating world.

------------

The results reported herein benefitted from participation in NASA's Nexus for
Exoplanet System Science (NExSS) research coordination network sponsored by NASA's
Science Mission Directorate.  

This work was also supported by NASA Goddard Space Flight Center ROCKE3D Science
Task Group funding.

Thanks to Linda Sohl and Jeff Jonas for useful discussion and help with the
Venus topographic overlay used in Figure 4. Thanks to Nancy Kiang for
her help in getting the 2.9Ga Solar Spectrum to work with the GISS 3D-GCM for
the paleo-Venus runs.

\end{abstracttext}

\end{document}